# The origin of Franson-type nonlocal correlation

Byoung S. Ham

School of Electrical Engineering and Computer Science, Gwangju Institute of Science and Technology,
123 Chumdangwagi-ro, Buk-gu, Gwangju 61005, S. Korea
(March 03, 2023)

**Abstract**: Franson-type nonlocal correlation is for the second-order intensity fringes measured between two remotely separated photons via coincidence detection, whereas their locally measured first-order intensities are uniform. This nonlocal intensity-product fringe shows a joint-phase relation of independent local parameters. Here, the Franson nonlocal correlation is investigated using a coherence approach based on the wave nature of quantum mechanics to understand the mysterious quantum feature of nonlocal fringes. For this, a typical Franson scheme based on entangled photon pairs is coherently analyzed for both local and nonlocal correlations, where the local intensities are due to many-wave interference between measured photos. For the nonlocal fringe, however, coincidence detection results in selective measurements, resulting in second-order amplitude superposition between locally measured photon's basis products. Due to the intrinsic property of a fixed sum-phase relation between entangled photons in each pair, the joint-phase relation of the nonlocal fringe is immune to the random spectral detuning of photon pairs. As in the first-order amplitude superposition of a single photon's self-interference, the second-order amplitude superposition between nonlocal basis-products is the origin of the nonlocal fringe.

**Introduction**
Quantum entanglement is known as a weird phenomenon that cannot be explained by classical physics or achieved by any classical means [1]. Ever since the well-known thought experiment by Einstein, Podolsky and Rosen (EPR) in 1935 [2], EPR has been a key aspect of quantum information science and technologies in computing [3-5], communications [6-9], and sensing areas [10-12]. Although Bell mathematically derived the so-called Bell inequality violation regarding the EPR paradox in 1964, the definition of classical physics rules out coherence optics [13]. Since then, mutual coherence (wave nature) between paired photons has not been carefully considered [13-26]. Here, a completely different coherence approach is applied for the Franson-type nonlocal correlation to understand otherwise mysterious nonlocal quantum feature. For this, we focus on the mutual coherence between paired photons generated from second-order nonlinear optics. This coherence interpretation on the observed nonlocal correlation fringe [20] gives us definite and clear reasoning-based physics.

Franson-type nonlocal correlation [19] has been studied since 1987 [20-26]. Unlike Bell inequality violations based on a polarization-basis control [13-18], Franson nonlocal correlation is for phase-basis control in a pair of unbalanced Mach-Zehnher interferometers (UMZIs). The path-length difference of each UMZI is designed with respect to interacting photons' effective coherence length, resulting in a complete incoherence feature. However, these uniform intensities result in nonlocal intensity-product fringes when coincidently measured. Although a mathematical analysis of the nonlocal correlation is clear for the nonlocal fringe formation [20], its physical understanding has been severely limited due to the local randomness. Thus, the fringe in Franson nonlocal correlation has been left as a mysterious quantum feature.

In the present paper, the origin of Franson-type nonlocal correlation is coherently investigated to clear out the weirdness of nonlocal fringes based on local randomness. For this, we analyze, firstly, the single photon-based UMZI in each party, where each pair of entangled photons is phase, frequency, or polarization correlated via a spontaneous parametric down conversion (SPDC) process [27-32]. The phase correlation between paired photons generated from the SPDC process is due to the phase matching condition of second-order nonlinear optics [27,32]. Secondly, the mysterious nonlocal intensity-product correlation fringe is directly derived from the local randomness in both UMZIs via coincidence detection. From this analysis, both solutions of local randomness and nonlocal fringes are coherently derived from coincidence detection-caused selective measurements. In the coherence approach, the coincidence detection-resulting second-order amplitude superposition is, however, still left as a mysterious quantum phenomenon, as in the first-order amplitude superposition of a single photon [33].

## Results

Figure 1(a) shows the original Franson scheme [20] using entangled photon pairs generated from the SPDC processes of Figs. 1(b) and (c) [32]. As studied [20-23] and applied for quantum key distributions [24-26], coincidence measurements between two remotely separated output photons show a path-length difference-dependent fringe in a joint-phase relation of local parameters, even though their local measurements do not. This nonlocal fringe for the second-order intensity correlation looks exactly the same as the first-order intensity correlation in a typical double-slit or MZI interference [33]. Regarding the UMZIs in Fig. 1(a) whose entangled photon source S is depicted in Figs. 1(b) and (c), both UMZIs are supposed to work as a coherence system to all individual photon pairs. The effective (ensemble) coherence of SPDC-generated photons in Fig. 1(b) is determined by the inverse of the spectral bandwidth $\Delta$, where $\Delta$ is inhomogeneously broadened by non-perfect phase-matching conditions. Thus, the coherence length of an individual pair should be much longer than the effective length $c\Delta^{-1}$, where c is the speed of light. Such a fact has already been experimentally demonstrated by using a bandpass filter [20].

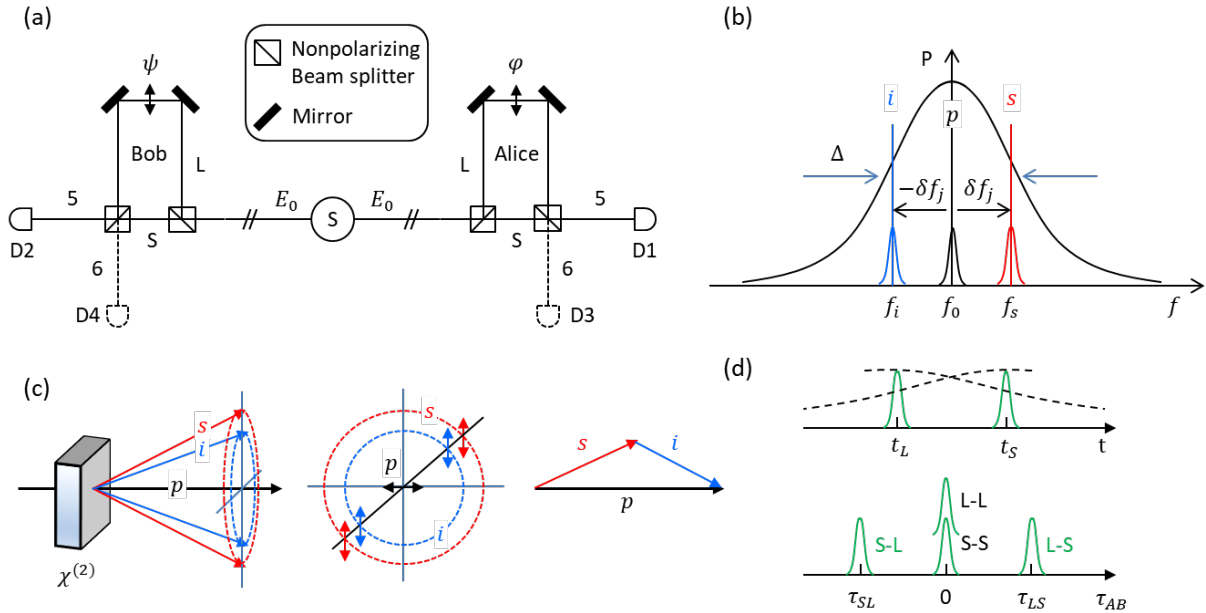

**Fig. 1. Schematic of Franson-type nonlocal correlation.** (a) The original Franson setup. (b) Probability distribution of the frequency of the light source S in (a). (c) Schematic of Type I SPDC. $2f_0 = f_s + f_i$. S (short path) and L (long path) indicate the two paths of each unbalanced MZI. (d) Schematic of coincidence detection. D: single photon detector. In (b), $\Delta$ is the full width at half maximum. In (b) and (c), $p$, $s$, and $i$ indicate pump, signal, and idler photons, respectively, where $s$ and $i$ can be swapped due to the phase matching condition (see text). $\tau_{SL}$ is the temporal delay between $\tau_S$ and $\tau_L$, where $\tau_i$ is the arrival time of a photon through path $i$. The dashed curves are for an individual photon's probability amplitude in the time domain.

By the energy conservation law, the pump laser's linewidth determines the sharpness (jitter) of the center frequency $f_0$ in Fig. 1(b), affecting the visibility of nonlocal fringes. Due to the phase matching of the $\chi^{(2)}$ process [27,31,32], the phase correlation must be satisfied among the pump ($p$), signal ($s$), and idler ($i$) photons in Fig. 1(b). Assuming a narrow linewidth of the pump laser, the sum phase of the signal and idler photons in each pair should be fixed with respect to the pump phase. Due to the random phase of the pump photon for different pairs by the spontaneous SPDC process, however, a random global phase should be applied to each entangled photon pair. Although this global phase does not affect local intensities due to the Born rule that a measurable quantity is the product of probability amplitude and its complex conjugate [34], the detuning-caused local phase ($\delta f_j T_{SL}$) plays a critical role to the mean intensity, where $T_{SL} (= T_L - T_S)$ is the photon's time delay between two paths of UMZI (analyzed below). Thus, it is a reasonable assumption that the path-length difference δL (L − S) of the UMZI can be set to be much shorter than the coherence length $l_c$ of an individual photon pair, but to be much longer than the effective coherence ($\Delta^{-1}c$), resulting in the critical UMZI condition $\Delta^{-1}c \ll \delta L \ll l_c$ for the present analysis.

Under this general understanding of the SPDC process, it is quite obvious that no interference fringe is observed for local intensities in both UMZI outputs due to $\langle \delta f_j T_{SL} \rangle \gg 1$. This incoherence feature is caused by the ensemble effect of all measured photons in a time domain. Regardless of the temporal resolution of detectors, $T_{SL}$ is long enough, so that $\delta f_j$-detuned photons are strongly affected. Such an ensemble-based incoherence feature is common in a Gaussian distributed atomic system or many-wave interference [35], whose individual coherence is much longer than the effective one. Either individual photon-based intensity sum in quantum optics or superposition-based interference in coherence optics, measured local intensities in the output ports of UMZI show the same incoherence feature of a uniform intensity. Regarding the nonlocal basis products, however, the coincidence detection is free from such an incoherence feature as analyzed below. Under this coincidence condition, a coherence solution of the nonlocal intensity fringe is induced to be sensitive to the path-length difference between UMZIs, which is immune to the random detuning $\delta f_j$.

Figure 1(c) shows schematic of the Type I SPDC process for the same polarization-based entangled photon pairs. Type II SPDC also works in the same way due to no polarization influence in both parties. Because of the energy and momentum conservation laws [30,31], the signal (*s*) and idler (*i*) photons can be chosen to be interchangeable in an ensemble, resulting in the entangled state, $|\psi\rangle = \frac{1}{\sqrt{2}}(|s\rangle_A|i\rangle_B + |i\rangle_A|s\rangle_B)$, where the subscripts A and B indicate Alice and Bob, respectively. For this analysis, taking either basis product or both in $|\psi\rangle$ makes no difference in the results.

Satisfying a critical condition $\Delta^{-1}c \ll \delta L \ll l_c$ in Fig. 1(a), a coherence approach for an input photon ($E_0$) results in the following matrix representation for the output photons of UMZI in Alice's side:

$$\begin{bmatrix} E_5 \\ E_6 \end{bmatrix}_{Alice} = [BS][V]_A[\varphi'][BS]\begin{bmatrix} E_0 \\ 0 \end{bmatrix}$$
$$= \frac{E_0}{2}e^{i\xi_j}\begin{bmatrix} |S\rangle_A + |L\rangle_A e^{i\varphi'_j} \\ i(|S\rangle_A - |L\rangle_A e^{i\varphi'_j}) \end{bmatrix}, \quad (1)$$

where $[BS] = \frac{1}{2}\begin{bmatrix} 1 & i \\ i & 1 \end{bmatrix}$, $[\varphi'_j] = \begin{bmatrix} 1 & 0 \\ 0 & -e^{i\varphi'_j} \end{bmatrix}$, $[V]_A = \begin{bmatrix} |S\rangle_A & 0 \\ 0 & |L\rangle_A \end{bmatrix}$, $\varphi'_j = \delta f_j T_{SL} + \varphi$, and $\varphi = f_j T^A_{PZT}$. $E_0$ is the amplitude of a single photon. Here, the common phase factor of the center frequency $f_0$ and a random global phase $\varphi^j_0$ to each entangled photon pair is taken out in terms of $\xi_j$, where $e^{i\xi_j}$ does not contribute to local intensities. Due to the energy conservation law in SPDC, the frequency detuning of the paired photons must be symmetric by $\pm \delta f_j$ from the center frequency $f_0$, where $f_0$ is half of the pump frequency, resulting in $f_j = f_0 \pm \delta f_j$ [31,32]. The term $f_j T^A_{PZT}$ is a local and independent control parameter, whose corresponding path length is in the order of the wavelength of photons: $T^A_{PZT}(\sim 10^{-6}) \ll T_{SL}(\sim 10^{-1})$. Considering $\Delta < 0.01 f_0$, thus, $\varphi = f_0 T^A_{PZT}$ is chosen for the coherence analysis. The $|S\rangle_A$ and $|L\rangle_A$ represent unit vectors of the short (S) and long (L) paths of the UMZI.

Likewise, the matrix representation for the UMZI at Bob's side in Fig. 1(a) is as follows:

$$\begin{bmatrix} E_5 \\ E_6 \end{bmatrix}_{Bob} = \frac{E_0}{2}e^{i\xi_j}\begin{bmatrix} |S\rangle_B + |L\rangle_B e^{i\psi'_j} \\ i(|S\rangle_B - |L\rangle_B e^{i\psi'_j}) \end{bmatrix}, \quad (2)$$

where $\psi'_j = -\delta f_j T_{SL} + f_j T^B_{PZT}$. Unlike the uncontrollable parameter $\delta f_j$, both $\varphi (= f_0 T^A_{PZT})$ and $\psi (= f_0 T^B_{PZT})$ are, however, controllable local parameters. Here, $\varphi$ and $\psi$ are of course independent and individual. From Eqs. (1) and (2), the amplitudes of the $j^{th}$ photon pair in all output ports in both sides are represented as follows:

$$E^A_{5j} = \frac{E_0}{2}e^{i\xi_j}(|S\rangle_A + |L\rangle_A e^{i\varphi'_j}), \quad (3)$$

$$E^A_{6j} = \frac{iE_0}{2}e^{i\xi_j}(|S\rangle_A - |L\rangle_A e^{i\varphi'_j}), \quad (4)$$

$$E^B_{5j} = \frac{E_0}{2}e^{i\xi_j}(|S\rangle_B + |L\rangle_B e^{i\psi'_j}), \quad (5)$$

$$E^B_{6j} = \frac{iE_0}{2}e^{i\xi_j}(|S\rangle_B - |L\rangle_B e^{i\psi'_j}). \quad (6)$$

In Eqs. (3)-(6), the superscript A (B) indicates Alice (Bob), and the subscripts 5 and 6 are for the output port numbers in both parties. The fixed short-path length difference between UMZIs can be included in either $\varphi$ or $\psi$ as a fixed phase (not shown). From Eqs. (3)-(6), thus, the corresponding mean intensities are obtained:

$$\langle I_{5j}^A \rangle = \langle \frac{I_0}{2}(1 + \langle S|L \rangle_A \cos\varphi_j') \rangle, \tag{7}$$

$$\langle I_{6j}^A \rangle = \langle \frac{I_0}{2}(1 - \langle S|L \rangle_A \cos\varphi_j') \rangle, \tag{8}$$

$$\langle I_{5j}^B \rangle = \langle \frac{I_0}{2}(1 + \langle S|L \rangle_B \cos\psi_j') \rangle, \tag{9}$$

$$\langle I_{6j}^B \rangle = \langle \frac{I_0}{2}(1 - \langle S|L \rangle_B \cos\psi_j') \rangle. \tag{10}$$

Here, the local coherence term of $\langle S|L \rangle_{A,B}$ plays a key role for the ensemble effects. Under the UMZI condition $\Delta^{-1}c \ll \delta L \ll l_c$, all mean local intensities become $\frac{I_0}{2}$ because of $\langle \cos\varphi_j' \rangle = \langle \cos\psi_j' \rangle = 0$, regardless of the local parameters, where $\delta f_j T_{SL}$ ($\gg 1$) is a dominant term. Even for a narrow-bandwidth $\Delta$, the uniform local intensities are satisfied unless $\Delta^{-1}c \ll \delta L$ is violated. As discussed below, this incoherence condition is quite important for the coincidence detection-caused selective measurements. Thus, the local randomness is coherently confirmed not by the particle nature but by the wave nature of an entangled photon pair.

We now analyze the general case of the Franson-type nonlocal correlation in Fig. 1. In both UMZIs, the two-photon basis-product can be represented by linear superposition of both paths S and L. Thus, there are four different path-basis products for nonlocal correlation measurements between space-like separated UMZIs. Firstly, the paired photons pass through only short paths in both UMZIs, resulting in a S-S relation. Secondly, the paired photons pass through only long paths, resulting in a L-L relation. Thirdly, the paired photons pass through either a short or long path alternatively, resulting in a S-L or L-S relation. Statistically, these four cases have equal probability amplitudes.

In Fig. 1(d), the top panel is for photons measured in each UMZI in an absolute time domain, while the bottom panel is for a coincidence-time domain ($\tau_{AB} = t_A - t_B$) between UMZIs. In the bottom panel, only the S-S and L-L cases are permitted for $\tau_{AB} = 0$ by definition of coincidence detection, whereas the S-L and L-S cases are prohibited due to $\Delta^{-1}c \ll \delta L$, resulting from timely and spatially separated wave packets. As a result, the role of coincidence detection is to remove distinguishable terms for S-L combinations caused by $\delta f_j T_{SL}$. Thus, the chosen two basis products at $\tau_{AB} = 0$ in the bottom panel of Fig. 1(d) are indistinguishable and can show second-order amplitude superposition, where the phase parameters are $\varphi_j' = f_0 T_{PZT}^A = \varphi$ and $\psi_j' = f_0 T_{PZT}^B = \psi$.

Using Eqs. (3) and (5), the amplitude $E_{AB}$ of coincidently measured nonlocal correlation $R_{AB}$ between two corresponding local measurements in Fig. 1(a) can be described for the indistinguishable basis products as:

$$E_{5j}^{AB}(\tau_{AB}) = \frac{I_0}{4} e^{2i\xi_j} \left(|S\rangle_A + |L\rangle_A e^{i\varphi_j'}\right)\left(|S\rangle_B + |L\rangle_B e^{i\psi_j'}\right)$$

$$= \frac{I_0}{4} e^{2i\xi_j} \left(|SS\rangle_{AB} + |LL\rangle_{AB} e^{i(\varphi_j' + \psi_j')}\right) \tag{11}$$

where both terms of $\langle L|S \rangle_{BA}$ and $\langle S|L \rangle_{BA}$ are ruled out by definition of coincidence detection as mentioned above. Here, $R_{5j}^{AB} = E_{5j}^{AB}(E_{5j}^{AB})^*$ and $E_{5j}^{AB} = E_{5j}^A E_{5j}^B$. Equation (11) is the second-order amplitude superposition corresponding to the first-order superposition in Eqs. (3)-(6). Here, indistinguishability between $|SS\rangle_{AB}$ and $|LL\rangle_{AB}$ is an essential condition for the quantum superposition between two nonlocal basis-products. This is the quintessence of the present coherence interpretation of the Franson-type nonlocal correlation in terms of the joint-phase relation, $\varphi_j' + \psi_j'$. As a result, the following nonlocal fringe is coherently achieved for the $j^{th}$ photon pair:

$$R_{5j}^{AB}(\tau_{AB}) = \frac{I_0^2}{8}(1 + \cos(\varphi_j' + \psi_j')), \tag{12}$$

where the first-order coherence between the short and long paths of the UMZI is a prerequisite for the cross basis-product terms $\langle SS|LL \rangle_{ABAB}$ and $\langle LL|SS \rangle_{ABAB}$. As in Eqs. (7)-(10) for the first-order amplitude superposition, these nonzero cross basis-product terms are the witness of the indistinguishable photon characteristics for the second-order amplitude superposition for the nonlocal fringe.

In Eq. (12), each random frequency detuning-caused joint phase term $\varphi'_j + \psi'_j$ is equal to $\varphi + \psi$ due to the opposite frequency detuning of the paired entangled photons in Fig. 1(b) satisfying the phase matching condition. Thus, the mean value of Eq. (12) becomes:

$$\langle R_{5j}^{AB}(\tau_{AB})\rangle = \frac{I_0^2}{8}(1 + \cos(\varphi + \psi)). \qquad (13)$$

Eq. (13) is immune to the random frequency detuning $\pm\delta f_j$ of both signal and idler photons in an ensemble. Equation (13) is also immune to the random global phase between entangled photon pairs due to individual and independent measurements. More importantly, Eq. (13) works even for $\tau_{AB} \neq 0$ until overall correlation between two wave packets diminishes in the order of $\tau_{AB} \sim \Delta^{-1}$ toward the classical lower bound. Thus, Eq. (13) is the coherence solution of the Franson-type nonlocal correlation [20]. For stabilized UMZIs in both parties, the mysterious nonlocal fringe in the Franson scheme is successfully understood from the coherence approach. Here, the indistinguishability between nonlocal basis products (S-S; L-L) in Eq. (11) is the new to the second-order quantum superposition between nonlocally measured photons via coincidence detection.

**Conclusion**
Franson-type nonlocal correlation was analyzed and discussed using the wave nature of photons for both local randomness and nonlocal correlation fringe. For this, a typical Franson scheme with an unbalanced MZI (UMZI) is investigated for entangled photons generated from a Type I SPDC process. In the present coherence approach, the UMZI is set to be coherent to individual photon pairs but incoherent to the ensemble of them. From this condition, firstly, the locally measured uniform intensity in each local detector was coherently confirmed to be uniform from a dephasing effect based on many-wave interference of spectrally broadened photons. Secondly, the nonlocal intensity correlation fringe was also coherently derived from coincidently detected local photons for the joint-phase relation satisfying inseparable basis products. For the joint-phase relation, the coincidence detection played an essential role to selectively choose a half of the tensor products between bipartite photons, resulting in the second-order amplitude superposition at the cost of 50 % event loss. Thus, the origin of nonlocal Franson correlation was completely understood using the coherence approach. In this coherence interpretation, the nonlocal coherence between locally measured photons was not contradictory to the violation of localism, where information is related with a wave packet caused by ensemble effects.

**Acknowledgement**
This work was supported by the ICT R&D program of MSIT/IITP (2023-2022-2021-0-01810), via Development of Elemental Technologies for Ultra-secure Quantum Internet.

**Author contribution**

**Conflict of interest**

**Data availability**

Data sharing not applicable – no new data generated.